\documentclass[twocolumn,showpacs,aps,prb,floatfix,eqsecnum]{revtex4}
\usepackage{amssymb,amsmath,hyperref,bm}
\usepackage{graphics}
\usepackage{graphicx}
\usepackage{psfrag,subfigure}

\begin{document}

\newcommand{\contraction}[5][1ex]{%
	\mathchoice
		{\contractionX\displaystyle{#2}{#3}{#4}{#5}{#1}}%
		{\contractionX\textstyle{#2}{#3}{#4}{#5}{#1}}%
		{\contractionX\scriptstyle{#2}{#3}{#4}{#5}{#1}}%
		{\contractionX\scriptscriptstyle{#2}{#3}{#4}{#5}{#1}}}%
\newcommand{\contractionX}[6]{%
	\setbox0=\hbox{$#1#2$}%
	\setbox2=\hbox{$#1#3$}%
	\setbox4=\hbox{$#1#4$}%
	\setbox6=\hbox{$#1#5$}%
	\dimen0=\wd2%
	\advance\dimen0 by \wd6%
	\divide\dimen0 by 2%
	\advance\dimen0 by \wd4%
	\vbox{%
		\hbox to 0pt{%
			\kern \wd0%
			\kern 0.5\wd2%
			\contractionXX{\dimen0}{#6}%
			\hss}%
		\vskip 0.2ex%
		\vskip\ht2}}
\newcommand{\contracted}[5][1ex]{%
	\contraction[#1]{#2}{#3}{#4}{#5}\ensuremath{#2#3#4#5}}
\newcommand{\contractionXX}[3][0.06em]{%
	\hbox{%
		\vrule width #1 height 0pt depth #3%
		\vrule width #2 height 0pt depth #1%
		\vrule width #1 height 0pt depth #3%
		\relax}}

\newcommand{\be}{\begin{equation}}
\newcommand{\ee}{\end{equation}}

\newcommand{\id}{\textrm{i}\delta}
\newcommand{\helta}{\hat{\Delta}}
\newcommand{\hamma}{\hat{\Gamma}}

\newcommand{\jsc}{J_\text{sc}}
\newcommand{\jcdw}{J_\text{cdw}}
\newcommand{\cs}{{\cal{C}}_s}
\newcommand{\cc}{{\cal{C}}_c}
\newcommand{\kf}{k_\text{F}}
\newcommand{\res}{\text{Res.}}

\newcommand{\im}{\textrm{i}}
\newcommand{\sgn}{\textrm{sgn}}
\newcommand{\bxi}{\bar{\xi}}
\newcommand{\lag}{{\cal L}}
\newcommand{\by}{ {\bf y} }
\newcommand{\bx}{ {\bf x} }

\newcommand{\xm}{ x_{-} }
\newcommand{\xp}{ x_{+} }
\newcommand{\ym}{ y_{-} }
\newcommand{\yp}{ y_{+} }
\newcommand{\mg}{{\mathrm g}}

\newcommand{\tr}{ \text{tr} }
\newcommand{\Tr}{ \text{Tr} }

\newcommand{\ev}[1]{\langle{#1}\rangle}
\newcommand{\vev}[1]{\langle 0|{#1}|0\rangle}

\newcommand{\up}{ \uparrow }
\newcommand{\dn}{ \downarrow }

\newcommand{\nn}{\hat n}
\newcommand{\bn}{{\bf N}}

\newcommand{\va}{{\vec a}}
\newcommand{\vo}{{\vec \Omega}}
\newcommand{\vs}{{\vec \sigma}}

\newcommand{\ttheta}{\tilde \theta}
\newcommand{\tphi}{\tilde \phi}

\newcommand{\pslash}{\ensuremath \raisebox{0.025cm}{\slash}\hspace{-0.2cm}\partial\/ }
\newcommand{\paslash}{\displaystyle{\not}\partial}

\def\LSCO{La$_{2-x}$Sr$_x$CuO$_4$}
\def\LBCO{La$_{2-x}$Ba$_x$CuO$_4$}
\def\YBCO{YBa$_2$Cu$_3$O$_{6+x}$}
\def\HBCO{HgBa$_2$Cu$_3$O$_{4+\delta}$}
\def\BKBO{BaKBiO}
\def\C60{A$_x$C$_{60}$}
\def\LNSCO{La$_{1.6-x}$Nd$_{0.4}$Sr$_x$CuO$_{4}$}
\def\optimalLCO{La$_{1.85}$Sr$_{.15}$CuO$_4$}
\def\VO{V$_2$O$_3$}
\def\TMTSF{(TMTSF)$_2$X}
\def\ET{BEDT...}
\def\hty{high temperature superconductivity}
\def\hts{high temperature superconductors}
\def\ie{ {\it i.e.\/} }
\def\eg{ {\it e.g.\/} }
\def\sign{ {\rm sign } }
\def\SRO{ Sr$_{1+n}$Ru$_{n}$O$_{3n+1}$}
\def\SROone{ Sr$_{2}$Ru$_{}$O$_{4}$}
\def\SROtwo{ Sr$_{3}$Ru$_{2}$O$_{7}$}
\def\SROinf{ Sr$_{1}$Ru$_{}$O$_{3}$}
\def\CRO{Ca$_{1+n}$Ru$_{n}$O$_{3n+1}$}
\def\LCO{La$_2$CuO$_4$}
\def\LCOplus{La$_2$CuO$_{4+\delta}$}
\def\LSNiO{La$_{2-x}$Sr$_x$NiO$_{4+\delta}$}
\def\BSCCO{Bi$_2$Sr$_2$CaCu$_2$O$_{8+\delta}$}
\def\oxychloride{Ca$_{2-x}$Na$_x$CuO$_2$Cl$_2$}
\def\LNSCO{La$_{1.6-x}$Nd$_{0.4}$Sr$_x$CuO$_{4}$}
\def\TMTSFClO{(TMTSF)$_2$ClO$_4$}
\def\HgCu3{HgCa$_2$Cu$_3$O$_{8+y}$}
\def\HgCu4{HgBa$_2$Ca$_3$Cu$_4$O$_{10+y}$}
\def\TlCu{Tl$_2$Ba$_2$CuO$_{6+\delta}$}
\def\TlCu3{Tl$_2$Ba$_2$Ca$_2$Cu$_3$O$_{10+y}$}
\def\TlCu4{Tl$_2$Ba$_2$Ca$_3$Cu$_4$O$_{12+y}$}
\def\TlCun{Tl$_2$Ba$_2$Ca$_{n-1}$Cu$_n$O$_{2n+4+y}$}
\def\HgCun{HgBa$_2$Ca$_{n-1}$Cu$_n$O$_{2n+2+y}$}
\def\BiCun{Bi$_2$Sr$_2$Ca$_{n-1}$Cu$_n$O$_y$}
\def\BiCu3{Bi$_2$Sr$_2$Ca$_{2}$Cu$_3$O$_y$}
\def\BiCaMnO{Bi$_{1-x}$Ca$_x$MnO$_3$}
\def\NCCO{Ne$_{2-x}$Ce$_x$CuO$_{4\pm\delta}$}
\def\8LSCO{La$_{1.88}$Sr$_{.12}$CuO$_4$}
\def\110LNSCO{La$_{1.5}$Nd$_{0.4}$Sr$_{0.1}$CuO$_{4}$}
\def\stage4LCO{La$_{2}$CuO$_{4+\delta}$}
\def\Y248{YBa$_2$Cu$_4$O$_8$}
\def\PCCO{Pr$_{2-x}$Ce$_x$CuO$_{4\pm\delta}$}
\def\HTS{High temperature superconductors}
\def\htr{high temperature superconductor}
\def\cdw{charge-density wave}
\def\cdws{charge-density waves}
\def\Cdws{Charge-density waves}
\def\NbSe2{NbSe$_2$}
\def\TaSe2{TaSe$_2$}
\def\TiSe2{TiSe$_2$}
\def\NaCoOH2O{Na$_{0.3}$CoO$_{2y}$H$_2$O}
\def\MgB2{MgB${}_2$}
\def\telephone{Sr$_{14-x}$Ca$_{x}$Cu$_{24}$O$_{41}$}


\title{Charge-Density-Wave and Superconductor Competition in Stripe Phases of High Temperature Superconductors}
\author{Akbar Jaefari}
\author{Siddhartha Lal}
\author{Eduardo Fradkin}
\affiliation{Department of Physics, University of Illinois at Urbana-Champaign, 1110 West Green Street, Urbana, Illinois 61801-3080, USA}

\date{\today}
\begin{abstract}
We discuss the problem of competition between a superconducting (SC) ordered state with a  charge density wave (CDW) state in stripe phases of high $T_c$ 
superconductors. We consider an effective model for each stripe motivated by studies of spin-gapped electronic ladder systems.
We analyze the problem of dimensional crossover arising from inter-stripe SC and CDW 
couplings using non-Abelian bosonization and renormalization group (RG) 
arguments to derive an effective $O(4)$-symmetric nonlinear $\sigma$-model in $D=2+1$ 
for the case of when both inter-stripe couplings are of equal magnitude as well as 
equally RG relevant. By studying the effects of various symmetry lowering perturbations, we determine the structure of the phase diagram and show that, in general, it 
has a broad regime in which both orders coexist. The quantum and thermal critical behavior is discussed in detail, and the phase coexistence region is found to end at 
associated
$T=0$ as well as $T>0$ tetracritical points. The possible role of hedgehog topological excitations 
of the theory is considered and argued to be RG irrelevant at the spatially anisotropic higher dimensional low-energy fixed point theory. 
Our results are also relevant to the 
case of competing N\'eel and valence bond solid (VBS) orders in quantum magnets 
on 2D isotropic square as well as rectangular lattices interacting via nearest 
neighbor Heisenberg exchange interactions.
\end{abstract}



\maketitle


\section{Introduction}
Over the past few years substantial evidence, both experimental and theoretical, has accumulated showing that the pseudogap regime of 
high $T_c$ superconductors (HTSC)  is related with to existence of charge and/or spin ordered states, stripe phases that break translation 
symmetry in the conducting layers, as well as uniform 
anisotropic nematic states (for reviews see see \cite{kivelson-2003,vojta-2009,fradkin-2010,fradkin-2010b} and references therein). In addition, 
homogeneous phases with ``hidden order'' involving some degree of time reversal symmetry breaking have also been proposed for the pseudogap 
phase \cite{chakravarty-2001c,varma-2006b}. An important question in this field is to determine if these ordered phases either compete with or are part of 
the mechanism of high $T_c$. 

 In principle, superconducting (SC) and charge/spin stripe orders  can coexist or compete. For instance,  in a well developed SC (gapped)
 coupled to a weak charge density wave (CDW), the CDW has little effect. Even deep in a $d$-wave SC, weak stripe order (a 
 unidirectional CDW) does not generally affect the low energy physics of the SC quasiparticles  as its wave vector generally does not span the 
 nodal points \cite{sachdev-2003,berg-2008}. Yet, in a weak coupling BCS approach to both orders, they usually (but not always) compete with each other, 
 and the enhancement of one type of order normally suppresses the other, and one has ``competing orders'' (see e.g. Ref.[\onlinecite{sachdev-2004}] and 
 references therein). 
However, recent work on stripe phases has provided a different perspective on this question by showing that in a strongly correlated system 
inhomogeneous phases not only are unavoidable \cite{kivelson-1998} but may also be part of the mechanism of SC and, moreover, 
that there exists an optimal 
degree of inhomogeneity at which the  $T_c$ is largest\cite{arrigoni-2004,kivelson-2007}.

In this work we consider the problem of the interplay between SC and stripe phases of HTSC and formulate an effective field theory of these competing/
coexisting orders.  We will use as a starting point a strong coupling picture of a stripe phase in which this 2D ordered state is represented as a quasi-1D 
system,  an array of doped Hubbard-type ladders  each in a Luther-Emery liquid, a phase with a finite, and typically large,
spin gap (denoted by $\Delta_s$) \cite{carlson-2000,arrigoni-2004}. There is extensive 
numerical evidence for this assumption based, primarily, on density matrix renormalization group (DMRG) calculations of Hubbard and $t-J$ models 
on up to seven leg ladders \cite{white-2000,carlson-2000,jackelmann-2005}. There is also supporting analytic evidence for spin gap phases in doped 
ladders\cite{wu-2003,tsvelik-2010} although the relation between the parameters of the effective (bosonized) field theory thus derived and microscopic 
models can only be reliably determined from numerics. In the low-energy regime these systems exhibit spin-charge separation. Furthermore, for a broad 
range of doping, $0\leq x \lesssim 0.3$, ladders with only repulsive interactions have a spin gap but do not have a charge gap (provided $x>0$) 
(for a review see \cite{carlson-2000}.) 
The states on the ladders represent the local high energy physics from which the low energy 2D states arise. (These models can also be used to describe 
ladder materials \cite{dagotto-1996}).  Thus, the resulting SC state is not the result of pairing in a system with preexisting quasiparticles. Instead, the 
quasiparticles of the SC state are emergent low energy excitations.

Based on this strong coupling picture of the stripe phase we use bosonization methods to determine the structure (and symmetries) of the effective low 
energy physics at scales well below the spin gap and construct an effective field theory for the resulting 2D state, a problem that was has not been addressed before. The power of this approach resides on the fact that 
allows the treatment of the competing effects of the tendency of the system  become a 2D superconductor  and a 2D charge ordered phase (a CDW). It 
turns out that, for a range of parameters, the system naturally has an approximate (enlarged) symmetry, that includes both the SC and stripe orders, that 
has far reaching consequences on the structure of the phase diagram. In particular, this approximate symmetry allows us to establish the existence of an 
intermediate phase in the 2D system in which CDW and SC orders coexist rather than being separated by a first order transition. This general question 
was discussed earlier on using inter-ladder renormalization group (RG) methods \cite{emery-2000,vishwanath-2001} and inter-chain mean field theory 
(ICMFT)\cite{carr-2002}, suggesting (in both cases) that the transition (at $T=0$) is first order and that there is an associated bicritical point. In this paper 
we will show that, instead, there is a phase in which SC and CDW coexist and that there is a tetracritical point (both at $T=0$ and at $T>0$.)

This work is is organized as follows. In Section \ref{sec:stripe} we present a summary of the stripe model and use it to determine the structure of the effective field 
theory. 
The effective field theory is derived in Section \ref{sec:effective}. The quantum and and thermal (classical) critical behaviors are discussed in Section \ref{sec:critical}. 
We close with a summary of results and extensions in Section \ref{sec:conclusions}.

\section{Analysis of the stripe model}
\label{sec:stripe}

 We consider an array of weakly coupled spin-gapped $1D$ systems with a gapless charge sector 
(``1CS0'')\cite{balents-1996}. We will follow the construction of Refs. [\onlinecite{emery-2000,arrigoni-2004}].
The low energy Hamiltonian for an array of decoupled ladders has the bosonized form
\be
	H_{\rm intra}= \sum_{i=1}^N\frac{v_c}{2} \int dx \left\{ K_c (\partial_x \theta_{c,i})^2 +\frac{1}{K_c} (\partial_x \phi_{c,i})^2  \right\} 
	\label{LL-hamiltonian}
\ee
where $i$ is the stripe label, $K_c$ is the charge Luttinger parameter of each ladder; $\phi_{i}$ is the phase  field of the CDW along each ladder and 
$\theta_{i}$ are the SC phase fields of each ladder. We will ignore other gapped degrees of freedom of the ladder as we are only interested in the physics below the spin gap $\Delta_s$.

The fields $\phi_i$ and $\theta_i$ are dual to each other, and satisfy the equal-time
commutation 
relations $[\phi_{i}(x), \partial_{x}\theta_{j}(x')]=i  \delta_{ij}\delta(x-x')$. The SC and CDW order parameters on each ladder in a spin gap phase are 
given by $\mathcal{O}_{SC}(x) \sim \exp(i \sqrt{2\pi}\theta(x))$ and $\mathcal{O}_{CDW}(x)\sim \exp(i \sqrt{2\pi}\phi(x))$, respectively. 

DMRG studies of hole-doped ladder systems~\cite{noack-1994} showed that $K_{c}$ is a function of doping $x$ with $K_c \to 2$ as $x \to 0$ to $K_c \to 1/2$ as $x \to x_c \simeq 0.3$.
In the presence of a spin gap the single particle hopping is irrelevant. The leading
relevant inter-stripe operators are the Josephson and CDW
interactions  \cite{emery-2000,vishwanath-2001,arrigoni-2004,kivelson-2007} 
\begin{align}
\begin{split}
	 H_\text{inter}&=\mathcal{J}_{SC} \sum_{\langle i,j \rangle} \int dx \cos\Big(\sqrt{2\pi} \left(\theta_{c,i}-\theta_{c,j}\right)\Big) \\
	 	& \quad + \mathcal{J}_{CDW} \sum_{\langle i,j \rangle} \int dx  \cos\Big(\sqrt{2\pi} \left(\phi_{c,i}-\phi_{c,j}\right)\Big)
\label{cdwscint}
\end{split}
\end{align}
Notice that we have used the seemingly wrong sign for the Josephson 
coupling (as it induces a $\pi$ phase shift between neighboring ladders). However this is the natural sign for the coupling between the CDW order parameters on neighboring ladders
since they want to couple with a $\pi$ phase shift so as to lower the effects of the repulsive interaction. We will refer to this as an ``antiferromagnetic'' sign.
However, since the two phase fields $\phi_j$ and $\theta_j$ are conjugate to each other (as required by their commutation relations), it is possible to carry out a 
unitary transformation on the SC phase fields $\theta_j$ on every other ladder (induced by the action of unitary operators involving only the CDW phase fields of 
the same ladder) and map the problem to one in which both interactions have a ``ferromagnetic'' 
sign, and both phases want to lock-in without a $\pi$ phase shift. Thus, in this simple state in which there is a spin gap on every ladder, the state associated with the sign alternation of a 
phase-density-wave\cite{berg-2007,berg-2009b} is equivalent to one without the phase alternation.   In contrast if both interactions had a ``ferromagnetic'' sign from the 
outset (i.e. if the phases wanted to lock 
without a $\pi$ phase shift)  it is no longer possible to map the system to one in which both are ``antiferromagnetic''. We will see below that this makes an important 
difference to the form of the effective field theory.

 In the spin gap regime, $0\leq x \leq x_C \simeq 0.3$,  the scaling dimensions of the CDW and Josephson couplings of Eq.\eqref{cdwscint} is $\Delta_{SC}=1/K_c$ and $\Delta_{CDW}=K_c$ respectively. Hence, both couplings are relevant since their scaling dimensions satisfy $\Delta_{SC,CDW} \leq 2$, with SC being more relevant than CDW for 
$x\lesssim 0.1$ (where $K_c=1$) and conversely for $0.1 \lesssim x \lesssim 0.3$ (where $\Delta_s \to 0$).  Then, a 
 perturbative RG analysis of these two inter-stripe interactions, valid for $\mathcal{J}_{SC}$ and $\mathcal{J}_{CDW}$ sufficiently small, shows that for 
 $0<x\lesssim 0.1$ the system flows to a 2D 
 (anisotropic) SC (with subdominant CDW correlations) whereas for $0.1 \lesssim x \lesssim 0.3$ it flows to an (anisotropic) crystal (a bidirectional CDW, with 
 subdominant SC correlations) suggesting a direct first order transition at $x \sim 0.1$ \cite{emery-2000,arrigoni-2004}.  We will only consider nearest-neighbor inter-stripe interactions, as couplings for further away stripes become small quickly  (although they are 
just as relevant.)

 Inter-chain mean field theory (ICMFT) \cite{scalapino-1975,carlson-2000,essler-2002} also found a direct SC-CDW first order transition\cite{carr-2002}. 
  ICMFT yields a good qualitative picture of the dimensional crossover and, indeed, it has the correct asymptotic scaling as $\mathcal{J}_{SC/CDW} \to 0$ (see, 
  e.g. Refs.[\onlinecite{affleck-1996,carlson-2000,arrigoni-2004}]), predicting the critical temperatures $T_{SC} \propto \mathcal{J}_{SC}^{\frac{K_c}{2K_c-1}}$, and 
  $T_{CDW} \propto \mathcal{J}_{CDW}^{1/(2-K_c)}$, respectively. Within ICMFT, close enough to the critical temperatures $T_{SC}$  and $T_{CDW}$, the order 
  parameter is 
  small and follows the critical behaviors~\cite{carr-2002,jaefari-2010}
$\Delta_{SC}(T)  \propto \Delta_{\text{SC}}(0) \left( 1-\frac{T}{T_{SC}} \right)^{ \frac{K_c}{4K_c-1}}$, and 
$\Delta_{CDW}(T) \propto \Delta_{CDW}(0) \left(1-\frac{T}{T_{CDW}}\right)^{1/(4-K_c)}$.
These results are consistent with those of Ref.[\onlinecite{carlson-2000}] for the case of $K_c=1/2$ (weakly coupled Luther-Emery stripes), which predicts a 2D SC 
phase with BCS-
type scaling. The problem of the competition of SC and CDW ordering was examined in considerable detail in Ref.[\onlinecite{carr-2002}] at the level of ICMFT, 
supplemented by a random phase approximation (RPA) analysis of fluctuation effects, and concluded that, at this level of approximation, there is a direct first 
order transition from SC to CDW order at $T=0$ .

However,  the problem of whether  in the 2D system there is a direct SC-CDW first order transition (and an associated bicritical point) or a phase in which both order 
parameters coexist (controlled by a tetracritical point) cannot be addressed at the level of ICMFT or within  the quasi-1D regime, and requires an RG analysis of the 
2D problem. 
In answering this question, we take an alternate route, and derive an effective field theory of the 2D system, and study it as a problem in (quantum and thermal) critical 
phenomena. 

In the regime where the SC/CDW competition is strongest, i.e. near the self-dual point (SDP) at $K_c=1$, 
both the SC and CDW interactions have the same scaling dimension  $\Delta_{SC}=\Delta_{CDW} =1$,
and hence must be treated on an equal footing.
Under duality, $\theta_i \leftrightarrow \phi_i$ and $K_c \leftrightarrow K_c^{-1}$, and
at $K_c=1$ the Hamiltonian $H=H_{\rm intra}+H_{\rm inter}$ of the 2D system is exactly self-dual (if the inter-stripe couplings are equal). Moreover, at $K_c=1$ the 
system  has a (dynamical) $SU(2)$ symmetry, a feature reproduced by ICMFT\cite{carr-2002}. 
The chiral fields on each ladder
$\phi_{R/L} = (\theta \pm \phi)/2$ can be used to define three chiral current operators 
\be
J_{R,L}^z \sim \partial_x \phi_{R,L}, \qquad
J_{R,L}^\pm \sim \exp\left( \pm i 2 \sqrt{2\pi}\phi_{R,L}\right)
\label{su_1}
\ee
 each with scaling dimension $1$, which generate an 
$SU(2)_1$ Kac-Moody algebra independently for the right and left moving degrees of freedom \cite{difrancesco-1997}. 
We will see below that at  $K_c=1$ $SU(2)_R\times SU(2)_L$  is also a symmetry of the full 2D Hamiltonian, Eqs. \eqref{LL-hamiltonian} and \eqref{cdwscint}, 
provided the 
inter-stripe couplings are equal. This symmetry is broken down to $U(1)_R\times U(1)_L$ both away from $K_c=1$ and by unequal couplings.
The total interacting Hamiltonian, $H=H_{\rm intra}+H_{\rm inter}$ is invariant under global $U(1)_R\times U(1)_L$ (chiral) symmetries. In the absence of 
inter-stripe couplings these global symmetries turn into sliding symmetries \cite{emery-2000,vishwanath-2001}.
We note that  the Hamiltonian in Eq.\eqref{LL-hamiltonian} at $K_c=1$ also represents the continuum limit of quantum Heisenberg antiferromagnetic chains. With some caveats discussed below, our results will also be relevant to the case of an array of such spin chains weakly coupled to one another.

\section{Effective Field Theory}
\label{sec:effective}

Let us now derive first an effective field theory in $2+1$ dimensions in the regime in which these SC and CDW  maximally compete 
with each other. Hence the effective field theory will have, to zeroth-order, an $SU(2)\times SU(2)$ symmetry that will be explicitly broken down to $U(1) \times U(1)$. 
In spirit, this approach is reminiscent to the phenomenological $SO(5)$ theory\cite{zhang-1997} although the  microscopic origin of the symmetry is quite different.
Provided $K_c\simeq 1$, it is possible to treat the symmetry breaking terms in a controllable fashion.  
To do this we begin by using the fact that, at $K_c=1$, the effective Hamiltonian of each ladder ($j=1, \ldots, N$), Eq.\eqref{LL-hamiltonian}, has a well known 
equivalent representation in terms of an $SU(2)_1$ Wess-Zumino-Witten model (WZW) whose 
Hamiltonian density is\cite{witten-1984,difrancesco-1997,gogolin-1998}
\be
\mathcal{H}_j=\frac{2\pi v_c}{3} \sum_{a=1}^3 \left(:J_{j,R}^a(x) J_{j,R}^a(x):+:J_{j,L}^a(x)J_{j,L}^a(x):\right)
\label{sugawara}
\ee
where the (normal ordered) products involve the right and left moving currents $J_{j,R}^a(x)$ and $J_{j,L}^a(x)$ ($a=1,2,3$), the three generators of two $SU(2)_1$ 
chiral Kac-Moody algebras. Away from $K_c=1$ there are symmetry-breaking ($SU(2)_R \times SU(2)_L \to U(1)_R \times U(1)_L$) intra-stripe couplings of the form 
\be
\mathcal{H}_{\rm intra}^{SB} \sim w \; \sum_j \left(:J_{j,R}^z(x) J_{j,L}^z(x):+:J_{j,L}^z(x)J_{j,R}^z(x):\right)
\ee
 with $w \to 0$ is $K_c \to 1$, as well as inter-stripe (generally anisotropic) current couplings\cite{emery-2000}. 
 As shown in Refs.[\onlinecite{emery-2000,vishwanath-2001}], sufficiently strong and long-ranged inter-stripe current-current interactions can turn the inter-stripe
SC and CDW interactions RG irrelevant and lead to the low-energy fixed point of the sliding TLL (or smectic liquid). We will focus here, however, with the
opposite case, i.e., that of dominant SC and CDW inter-stripe interactions.

The WZW model is a non-linear sigma model (NLSM) in $1+1$ dimensions whose degree of freedom is a field $g(x)$ that takes values on a (compact) Lie group, 
$SU(2)$ in this case. The action of the WZW models for each stripe is \cite{witten-1984}
\begin{align}
	S^{k}_\text{WZW}[g] & = \frac{k}{16\pi} \int  d^2 x \,\, \tr \left( \partial^\mu g^\dagger \partial_\mu g \right) \nonumber \\
	& - \frac{k}{24\pi} \int_B d^3x \epsilon^{\alpha\beta\gamma} \tr \left( g^\dagger \partial_\alpha g \, g^\dagger \partial_\beta g \, g^\dagger \partial_\gamma g \right)
\label{wzwaction}
\end{align}
where $k$ is the level of the Kac-Moody algebra; here we are interested in the case $SU(2)_1$ and $k=1$. The second term in Eq.\eqref{wzwaction} is the  WZW 
term, where $B$ denotes a 3D solid sphere whose boundary $S^2$ is $1+1$ dimensional space-time. The $SU(2)$-field $g$ and $(\phi,\theta)$ fields are related by 
\be
	g_{\sigma\sigma'} \sim
		\begin{pmatrix}
			e^{-i \sqrt{2\pi}\phi} & -e^{i \sqrt{2\pi}\theta}\\
			e^{-i \sqrt{2\pi} \theta} & e^{i \sqrt{2\pi}\phi}
		\end{pmatrix}
\ee
The (combined) inter-stripe terms in the Hamiltonian density now take the form
\begin{eqnarray}
\mathcal{H}_{\rm inter}&=&-\mathcal{J}_+ \sum_j \textrm{tr}\big(g^{\dagger}_j(x) g_{j+1}(x)\big)\nonumber \\
&-&\mathcal{J}_- \sum_j \textrm{tr}\big(g^{\dagger}_j(x) \sigma_z g_{j+1}(x)\sigma_z\big)
\label{Hinter-su2}
\end{eqnarray}
where $\sigma_z$ is the diagonal Pauli matrix, and we have defined the couplings $\mathcal{J}_\pm=\mathcal{J}_{SC}\pm \mathcal{J}_{CDW}$. 
Notice that for $\mathcal{J}_-=0$ the $2+1$-dimensional system still enjoys an $SU(2)\times SU(2)$ symmetry, which is broken down to $U(1) \times U(1)$ 
if $\mathcal{J}_-\neq 0$. In addition, away from $K_c=1$, the intra-stripe Hamiltonian has additional terms of the form 
$\textrm{tr} (\partial_\mu g^\dagger \sigma_z \partial^\mu g \sigma_z)$ that also break the symmetry down to $U(1) \times U(1)$.

Depending on the signs of  $\mathcal{J}_\pm$, the inter-stripe couplings favor ordered phases in $2+1$ dimensions in which the $SU(2)$-valued matrix field 
$g$ is either uniform across the system or changes {\em sign} (i.e be staggered by an element of the $\mathbb{Z}_2$ center of the group $SU(2)$) from one stripe to 
the 
next. In a (relatively) recent paper, Senthil and Fisher \cite{senthil-2006} discussed the behavior of an array of antiferromagnetic quantum Heisenberg model in the 
quasi-1D regime and proposed a description of that system which is similar in spirit to the one we use here for the stripe state in the spin gap regime. One important 
difference is that in the Senthil-Fisher the $SU(2)$ group is associated with the spins which, in the 1D limit, have quasi-long range antiferromagnetic commensurate 
order with wave vector $\pi$. As a result, in Ref.[\onlinecite{senthil-2006}] the interchain coupling has the form $\textrm{tr} (g_i g_{i+1})+\textrm{h.c.}$ that breaks the 
symmetry to a single global $SU(2)$ symmetry. This coupling favors a 
staggered order in which $g$ alternates with $g^\dagger$ between neighboring chains. In the problem we discuss here this form of coupling is not allowed by gauge 
invariance (in the case of the SC order) and by translation invariance (in the case of CDW order). As we will see below this leads to some differences in the 
structure of the effective field theory.

The derivation of the effective field theory in $2+1$ dimensions proceeds in two stages, e.g. along the lines discussed early on by 
Affleck and Halperin\cite{affleck-1996}. In the quasi-1D limit the leading relevant operators are the inter-stripe couplings shown in Eq.\eqref{Hinter-su2}. In the 1D limit 
at the $SU(2)$ invariant system (with $K_c=1$), in the absence of the inter-stripe interactions, the low energy physics is controlled by 
the WZW fixed point\cite{witten-1984}, an infrared stable fixed point at a finite value of the NLSM coupling constant. 
At the WZW fixed point the inter stripe operators have scaling dimension $1$, 
strongly destabilize the fixed point of the WZW model, and 
favor $2+1$-dimensional ordered states at which the field $g$ picks up a non-vanishing expectation value.  In the presence of the ``internal anisotropy'' 
terms that break 
$SU(2)\times SU(2) \to U(1) \times U(1)$ the resulting ordered phases correspond to 2D SC, 2D CDW and possibly coexistence phases (which is the main question 
we address here). 

After this initial step of quasi-1D renormalization, the system becomes coarse-grained and flows to a full $2+1$-dimensional theory with the proper 
symmetries, with an effective field theory of the form of a relativistic-like $2+1$-dimensional NLSM. The effective theory is generally spatially anisotropic and also has 
terms ``internal'' anisotropy terms. However, the infrared unstable (quantum critical) fixed point of the $2+1$-dimensional NLSM occurs at a {\em finite} value of the 
NLSM effective coupling constant and, hence, cannot be accessed perturbatively from the quasi-1D regime (at least not in a controllable way). Thus, derivations of the 
effective theory based on naive gradient expansions of the quasi-1D Hamiltonian must be regarded as being only suggestive (at best) of the structure of the effective 
field theory near the quantum 
critical point. Nevertheless it is possible to use the powerful constraints of locality and symmetry to write down the structure of the effective field theory. This is the 
approach we use here. With one 
significant caveat concerning  the role of topology (that we will discuss below), the spatial anisotropies discussed above 
lead to redundant operators whose effects  can be taken into account by a suitable rescaling of the spatial coordinates and time. In contrast, the  
internal anisotropy terms play a key role. In what follows we will work in imaginary time. 

For reasons of clarity we find it useful to represent the $SU(2)$-valued matrix field $g$ in terms of a four-component vector $N^a$ ($a=0,1,2,3$) that 
takes values on the three-sphere $S^3$, i.e. $N^a=\frac{1}{2} \textrm{tr} (g \sigma^a)$, where we have used the basis of Hermitian $2 \times 2$ matrices, $\sigma^0=I$ 
and the three Pauli matrices for $a=1,2,3$. The four-component field $N^a$ satisfies the constraint $N^2=1$, and as 
such takes values on $S^3$.

Given these symmetry requirements, the only allowed effective action of the effective field theory in $2+1$ dimensions  is that of an $O(4)$ NLSM. 
Ignoring for the moment 
spatial anisotropies and internal anisotropy terms, the effective Lagrangian is
\be
	  \mathcal{L}_{0}[N] = \frac{1}{2g_0}  (\partial_\mu N)^2 
	  \label{nlsm-2+1}
	  \ee
	  where $\mu=0,1,2$ label time and the two spatial coordinates respectively.
The bare value of the effective coupling constant $g_0$ of the NLSM in $2+1$ dimensions has units of (length)$^{-1}$. Its value is essentially given by the geometric 
mean of the 
(suitably dimensionalized) coupling constant of the WZW model, $1/(4\pi a_0)$ (where $a_0$ is the stripe spacing) and the  inter-stripe coupling $\mathcal{J}_+$.  This value of $g_0$ will be substantially modified by renormalization effects.

The competition/coexistence of SC and CDW orders can be now be discussed by considering the effects of  operators 
that break the large $SU(2)\times SU(2)$ symmetry down to $U(1)\times U(1)$. The symmetry lowering arises from (i) a finite asymmetry in the bare 
values of the SC and CDW couplings, $\mathcal{J}_{SC/CDW} \neq 0$, and (ii) a departure from the SDP as $K_c-1\neq 0$. Up to spatially anisotropic 
gradient terms that we will omit for now, we obtain  the following effective Lagrangian 
\begin{eqnarray}
	  && \!\!\!\!\!\!\!\!\! \!\!\! \mathcal{L}_{\rm eff}[N] =\frac{1}{2g_0}(\partial_\mu N)^2 - w\; \partial_\mu N_a O_{ab} \partial_\mu N_b  - {\tilde h}\; N_a O_{ab} N_b 
	  \nonumber\\
	  &&
	  \label{O-not-same}
\end{eqnarray}
where $a,b=0,1,2,3$. In Eq.\eqref{O-not-same} the second term with coupling constant $w \sim K_c-1$ parametrizes the departure from the SDP, the third term with 
coupling constant 
$\tilde h \sim \mathcal{J}_-$ describes the unequal couplings between CDW and SC order parameters, and
$O$ is a fixed matrix that breaks $SU(2) \times SU(2) \to U(1)\times U(1)$.

There is still one more operator that can be part of the effective action that needs to be considered: the topological charge
$Q_\text{top}$ defined by
\be
Q_{\rm top}[N]=\frac{1}{12\pi^2} \int d^3x \epsilon_{\mu \nu \lambda} \epsilon_{abcd} N_a \partial_\mu N_b \partial_\nu N_c \partial_\lambda N_d
\label{Qtop-O(4)}
\ee
$Q_{\rm top}[N]$ is an integer, a topological invariant that classifies the non-trivial maps of the $2+1$-dimensional Euclidean space-time (compactified to $S^3$) to 
the target manifold of the 
$O(4)$ NLSM which is also isomorphic to $S^3$. In the language of homotopy theory these maps are classified by $\Pi_3(S^3) \simeq \mathbb{Z}$. Since the system 
at hand is time-reversal invariant, the only allowed topological term in the effective (Euclidean) action must have the form $S_{\rm top}=i p \pi Q_{\rm top}[N]$, where 
$p \in \mathbb{Z}$. Since $Q_{\rm top}[N]$ is also an integer, such a topological term has an effect only for $p$ odd.

\section{Classical and Quantum Critical Behavior} 
\label{sec:critical}

\subsection{$T=0$}

We begin the discussion of quantum criticality by considering the role of the topological term. 
Terms of this type play a crucial role in 1D spin-$1/2$ quantum Heisenberg 
antiferromagnets \cite{haldane-1983}, also described by an $SU(2)_1$ WZW 
model\cite{affleck-1986}. Extensions to higher dimensional antiferromagnets have also been 
proposed before and examined in some detail\cite{fradkin-1991,abanov-2000,tanaka-2005,senthil-2006}, 
but never in the context of stripe phases in superconducting systems. Particularly relevant here is the work of Senthil and 
Fisher\cite{senthil-2006} who discussed this same effective field theory in the context of the quantum phase transition from a 2D N\'eel 
antiferromagnetic state to a four-fold degenerate valence bond solid (VBS) state on a square lattice, whic has been conjectured to be controlled by a deconfined 
quantum critical (DQC)
point \cite{senthil-2004}. 
The topological excitations that carry a non-trivial winding number $Q[N]$ are monopole (``hedgehog'') configurations in 3D Euclidean space-
time, the instantons of this theory. Contrary to their 2D cousins, the Euclidean action of instantons  of  3D NLSMs is linearly divergent and 
hence are suppressed throughout the ordered phase. 
Nevertheless, within the DQC scenario they are still argued to play a key role both at the quantum critical point and in the quantum disordered phase (which 
becomes a topological phase). However, the analysis of Ref.[\onlinecite{senthil-2006}] shows that the 3D DQC fixed point is unstable to the effects of {\em spatially 
anisotropic} perturbations, such as the ones we have in this theory, and becomes inaccessible. The resulting effective field theory of our system is then the anisotropic 
NLSM of Eq.\eqref{O-not-same} but {\em without} the topological term which becomes an irrelevant operator at the accessible fixed point. 

\begin{figure}[hbt]
\begin{center}
\includegraphics[width=0.4\textwidth]{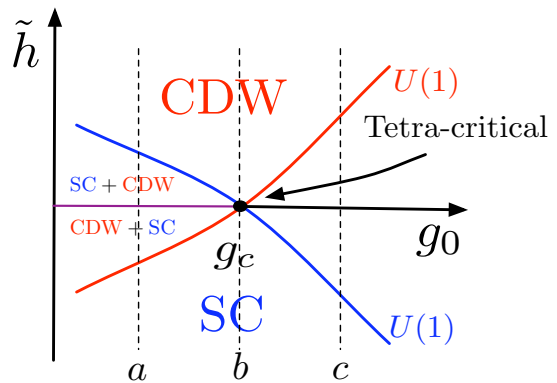}
\end{center}
\caption{(color online) Schematic phase diagram for a 2D stripe phase at $T=0$: $g_0$ is the coupling constant of the $O(4)$ NLSM and $\tilde h$ is the anisotropy. 
Here we assumed that the tetracritical point is a decoupled fixed point. (See text for details.)}
\label{phase-T0}
\end{figure}

Having determined the form of the effective field theory we can proceed to study its quantum and classical critical behaviors. 
The analysis of the phase diagram can now be determined using well established methods of (classical and 
quantum) critical phenomena and, in particular, of multicritical phenomena. At this new, more conventional, fixed point spatial anisotropies become mild 
redundant operators and their presence no longer  affect the critical behavior.
We will see now that the effective action of Eq.\eqref{O-not-same}  will allow us to find a answer to the problem of SC/CDW competition vs. coexistence, both at $T=0$ 
and at $T>0$. A summary of the results is presented in the phase diagrams shown in Fig.\ref{phase-T0} and Fig.\ref{phase-T>0}.

We thus have a $2+1$-dimensional NLSM with symmetry breaking fields. 
By power counting we see that the third term in Eq.\eqref{O-not-same} is the most relevant perturbation. From this point of view the problem of the CDW/SC 
competition is conceptually similar to other problem in which there is a partial breaking of the internal 
symmetry, e.g. the spin flop transition in magnets. This point of view will give us the solution of the problem. The critical behavior obeyed by this system must be 
approached either by means of a) a $2+\epsilon$ expansion of the NLSM (here $2$ means $1+1$ Euclidean space-time dimensions), 
or b) a $4-\epsilon$ expansion of the 
associated Landau-Ginzburg-Wilson (LGW) theory, usually known as $\phi^4$ field theory (again, here $4$ means $3+1$ Euclidean space-time dimensions). 
Although we already have the problem expressed as a NLSM we will have to use the $4-\epsilon$ expansion approach. The reason is that the $4-\epsilon$ 
expansion of the $\phi^4$ has a property known as  Borel-summability which allows for an accurate determination of its critical exponents 
directly in $D=3$ Euclidean space-time dimensions (see, e.g. Ref.[\onlinecite{zinn-book}]). In contrast, the conceptually simpler $2+\epsilon$ approach has poor 
convergence properties and has never been successfully used to compute exponents in $D=3$ even for the simplest NLSM. 

For this reason we will replace the NLSM 
effective action for the field $N$ with $O(4)$ symmetry by a theory with the same symmetries but in which the constraint of the NLSM is replaced by a suitable 
potential. Let us denote by $N_\phi$ the upper two (CDW) components  and $N_\theta$ the lower two (SC) components of the four-component scalar field 
 $N$, respectively. In this form we are describing the breaking of $O(4) \to O(2) \times O(2)$. This is a particular case of the breaking of $O(n)$ down to 
 $O(n_1)\times O(n_2)$, with $n=n_1+n_2$, that has been studied in detail in the literature. It was originally studied to one-loop order in the $4-\epsilon$ expansion by 
 Kosterlitz et al \cite{kosterlitz-1976}, and was more recently reexamined by Calabrese et al \cite{calabrese-2003} who used a five loop $4-\epsilon$ expansion with 
 Borel resummation and were able to determine the critical behavior in $D=3$ with high precision. 

The resulting LGW Lagrangian (or ``free energy density'') has the form
\begin{eqnarray}
&&\mathcal{L}[\vec N_\phi,\vec N_\theta]=\frac{1}{2} \left(\partial_\mu \vec N_\phi \right)^2+\frac{1}{2} \left(\partial_\mu \vec N_\theta \right)^2+\frac{r_\phi}{2}  \vec 
N_\phi^2+\frac{r_\theta}{2} \vec N_\theta^2
\nonumber \\
&&+\frac{u_\phi}{4!} \left(\vec N_\phi^2\right)^2+ \frac{u_\theta}{4!} \left(\vec N_\theta^2\right)^2+\frac{2w}{4!} \vec N_\phi^2 \vec N_\theta^2 \nonumber\\
&&
\end{eqnarray}
As usual, $r_\phi$ and $r_\theta$ measure the departure from the (classical or mean field) critical point; $r=(r_\theta+r_\phi)/2$ qualitatively plays the role of the 
coupling constant $g_0$ of the NLSM, $\tilde h=(r_\theta-r_\phi)/2$ of the symmetry breaking field, and $u_\theta$, $u_\phi$ and $w$ are four coupling constants that 
parametrize the potential. For $u=u_\theta=u_\phi=w$ the quartic terms have $O(n)$ symmetry.
Notice also that in this form the spatial anisotropies can always be absorbed by a suitable rescalings of the coordinates and fields.

Below $4$ dimensions, the free field (Gaussian) fixed point is always unstable.  
Kosterlitz et al\cite{kosterlitz-1976} showed that theories of these type can describe either a bicritical point (the endpoint of a line of direct first order transitions 
between the two phases with order parameters $\vec N_\phi$ and $\vec N_\theta$) or a tetracritical point (the endpoint of a region of phase coexistence of the two 
order parameters). They also showed that when the bicritical scenario holds, the fixed point associated with the critical endpoint has maximal symmetry, $O(n)$. In the 
tetracritical scenario, the $O(n)$ fixed point is unstable and two possibilities arise for the endpoint: a) either it is a decoupled fixed point (DFP) at which the 
$O(n_1)\times O(n_2)$ theory is effectively decoupled ($w\to 0$), or b) its is a biconical fixed point (BFP) at which the $O(n_1)\times O(n_2)$ has a 
non-trivial coupling $w^*$. 

The one-loop analysis of Kosterlitz et al\cite{kosterlitz-1976} predicts a tetracritical point with a DFP for some sufficiently large value of $n>4$, and a bicritical point for 
the spin-flop transition $O(3) \to O(2) \times \mathbb{Z}_2$. However, the one-loop results are unable to resolve the case of interest here, $O(4) \to O(2) \times O(2)$. 
On the other hand, the five-loop results of Calabrese et al\cite{calabrese-2003} show without ambiguity that in $D=3$ the $O(4) \to O(2) \times O(2)$ theory has 
a tetracritical point and not a bicritical point as the $O(4)$ invariant fixed point is unstable. However, the five-loop results do not have sufficient accuracy to distinguish 
between a tetracritical point controlled by a DFP or by a BFP. 
Nevertheless, regardless of this technical issue, this analysis predicts the existence of a phase in which the SC and CDW order parameters coexist at $T=0$. This is 
shown in the schematic phase diagram of Fig.\ref{phase-T0} as the phase labeled by SC+CDW. The relative strength of the SC and CDW order parameters varies 
continuously across this phase as the magnitude (and sign) of the coupling $\tilde h$ to the symmetry breaking field varies. (The broken line in Fig.\ref{phase-T0} is 
not a phase transition and denotes the manifold with higher 
symmetry). All the phase boundaries are in the universality class of the 3D classical XY model (i.e. the $O(2)$ 
Wilson-Fisher fixed point in $D=3$). Whether the tetracritical point is governed by a different coupled fixed point (BFP) or a decoupled one (DFP) is not 
presently established.

We end this subsection by discussing briefly the extension of this analysis to a stripe phase in 3D. For simplicity we will assume that the stripe phase consists of a 
stack of 2D stripe phases and, hence, that the order parameter theory is the same as in 2D. Much of the analysis above here follows here too. The main difference is 
that the effective field theory is now in $D=4$ Euclidean space-time dimensions. In this case the $O(4)$ NLSM does not have a topological term to begin with. In 
$D=4$ Euclidean space-time dimensions the free-field (Gaussian) fixed point is marginally stable and the classical (mean field) results are correct  up to logarithmic 
corrections to scaling. In this case there is clearly a phase in which SC and CDW orders coexist. The resulting phase diagram in 3D at $T=0$ has the same topology 
as in Fig.\ref{phase-T0}.

\subsection{$T>0$}

\begin{figure}[hbt]
\begin{center}
\subfigure[ \; $T>0, \; g_0>g_c$]{\includegraphics[width=0.4\textwidth]{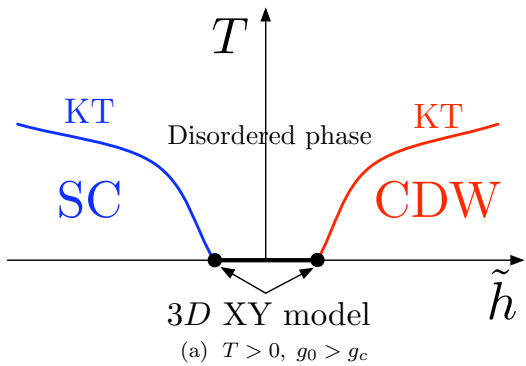}}
\subfigure[ \; $T>0, \; g_0=g_c$]{\includegraphics[width=0.4\textwidth]{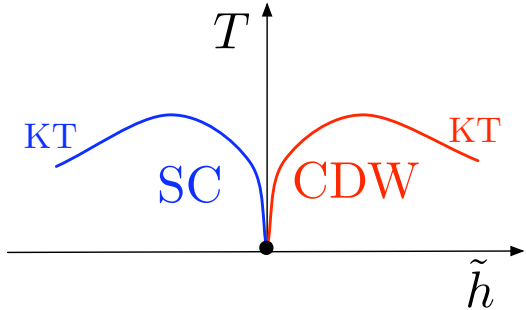}}
\subfigure[\; $T>0, \; g_0<g_c$]{\includegraphics[width=0.4\textwidth]{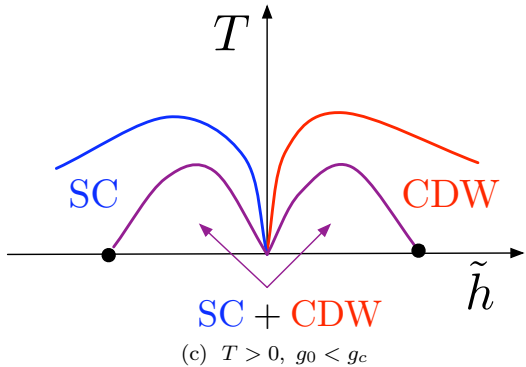}}
\end{center}
\caption{(color online) Schematic $T>0$ phase diagrams for a 2D stripe state as a function of the  anisotropy $\tilde h$ of the $O(4)$ NLSM for values of the coupling constant $g_0$ corresponding to the lines labelled by a, b and c in Fig.\ref{phase-T0}. See text for details.
}
\label{phase-T>0}
\end{figure}

\begin{figure}[hbt]
\begin{center}
\includegraphics[width=0.4\textwidth]{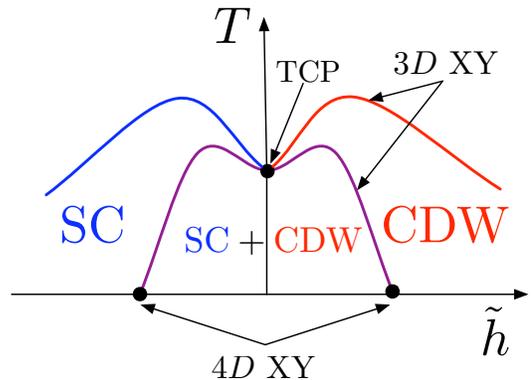}
\end{center}
\caption{(color online) Schematic $T>0$ phase diagrams for a 3D stripe state as a function of the  anisotropy $\tilde h$ of the $O(4)$ NLSM for values of the coupling constant $g_0<g_c$. The thermal transitions are 3D XY and the $T=0$ transitions are 4D XY. TCP labels the tetracritical point. See text for details.
}
\label{4Dphase-T>0}
\end{figure}

We now turn to the phase diagram at finite temperature, $T>0$.  It is not possible to replace the NLSM by the LGW theory in 2D at finite temperature,
since in 2D there is a drastic difference in the thermal critical behavior of the system if it is fine-tuned to have the larger $O(4)$ symmetry from where it has a 
lower $O(2) \times O(2)$ symmetry. (Recall that topological terms do not play a role in the thermal critical behavior as they always contain time derivatives of the 
fields.) (This was already emphasized by Carr and Tsvelik\cite{carr-2002}.) In both cases there is no long range order in 2D as required 
by 
the Mermin-Wagner theorem. However, the critical temperature of the $O(4)$-invariant  classical 2D NLSM is zero and it is in a classically disordered phase at all 
$T>0$, as shown in the schematic phase diagrams of Fig.\ref{phase-T>0} (a-c).

Away from the $O(4)$ symmetric theory, i.e. for $\tilde h \neq 0$, there are phase transitions at finite $T>0$. Since the symmetry is now reduced to $O(2) \times O(2)$ 
these are, generically Kosterlitz-Thouless (KT) transitions. At a fixed value of the NLSM coupling $g_0$ and anisotropy $\tilde h$ different sequences of phase 
transitions take place. Some details of the phase diagram depends on whether the $T=0$ quantum phase transition is described by a DFP or a BFP. In general there 
are three situations, shown in the phase diagrams of Fig.\ref{phase-T>0} (a-c). 

In Fig. \ref{phase-T>0} (a) we depict the case in which at $T=0$ the isotropic $O(4)$ NLSM is quantum disordered, $g_0>g_c$, corresponding to line $c$ in the $T=0$ 
phase diagram of Fig.\ref{phase-T0}. For some range of anisotropy $\tilde h$ the system remains disordered at all temperatures. 
At some critical anisotropy $\tilde h(g_0)$ there is a $T=0$ quantum phase transition to a SC or a CDW phase (depending on the sign of $\tilde h$), which is in the 
universality of the 3D XY classical model. For $T>0$ this transition becomes a KT transition and the SC (or the CDW phase depending of the case) has power law 
correlations and not long range order. Here we assumed that either the $T=0$ tetracritical point is a DFP or, in the case of of a BFP, that $g_0$ is above the region of 
the SC+CDW coexistence phase.

In Fig.\ref{phase-T>0} (b) we depict  the case in which $g_0=g_c$ (the QCP of the $O(4)$ NLSM), line $b$ in Fig.\ref{phase-T0}. $T_c=0$ only for  $\tilde h=0$. The 
situation depicted in the figure assumes that the $T=0$ tetracritical point is again of the DFP type and hence, that as $\tilde h$ increases (in magnitude) one does not 
enter the coexistence phase, but instead the SC phase (or CDW depending of the case) which undergoes a KT transition to the high temperature phase.

Finally, in Fig.\ref{phase-T>0} (c) we have the case $g_0<g_c$, line $a$ in Fig.\ref{phase-T0}. Once again we have $T_c=0$ at $\tilde h=0$ as we are in 2D. As a 
function of $\tilde h$ (at fixed $T$) one enters first the SC (or CDW) phase (with power law correlations, once again). There is a range of $\tilde h$ in which the 
coexistence phase appears (with KT transitions at both ends) and eventually a KT transition to the high temperature phase.

Let us finally discuss, briefly, the finite temperature behavior of the 3D stripe state. The main change compared to the 2D case is that the $O(4)$ symmetric NLSM has now a finite temperature transition, in the universality class of the 3D $O(4)$ Wilson-Fisher fixed point. Hence, although $T_c$ for the $O(4)$ NLSM is lower than that of the system with a lower $O(2) \times O(2)$ symmetry, it is not suppressed down to $T=0$ as in 2D. The 3D case has a phase diagram with a more conventionally looking tetracritical point.

\section{Discussion}
\label{sec:conclusions}

In this paper we have considered the problem of competition and coexistence of SC and CDW orders in stripe models of strongly correlated systems (e.g. weakly 
coupled ladders in the spin gap regime). We have shown that the natural effective field theory in the $2+1$-dimensional regime is a spatially anisotropic $O(4)$ 
NLSM with additional interactions that break the symmetry down to $U(1) \times U(1)$. We examined the quantum and thermal critical behaviors of this system. In the 
quantum regime we used the $\phi^4$ version of this theory (i.e. the Landau-Ginzburg-Wilson theory) with $O(4)$ global symmetry explicitly broken down to 
$U(1) \times U(1)$. Using relatively recent high precision five loop results (improved with Borel-Pad\'e resummation of the $\epsilon$ expansion) by 
Calabrese et al\cite{calabrese-2003} we showed that there is a phase in which SC and CDW orders coexist and that this phase transition is controlled by a tetra-
critical point (likely of the decoupled type). Our results for the quantum and thermal phase diagrams (and phase transitions) are summarized in the 
Figures \ref{phase-T0}, \ref{phase-T>0}, and \ref{4Dphase-T>0}.  It is worth noting that 
our RG phase diagrams are essentially the same irrespective of the sign of the CDW and SC couplings: we can map one 
problem onto the other by translating the $(\theta_{i},\phi_{i})$ fields in equation (\ref{cdwscint}) by 
$\pi$ on all odd-numbered stripes. Thus this analysis also describes quantum phase transitions in a spin-gapped version of the pair-density-wave (PDW) state, a 
phase in which charge and SC orders are intertwined.\cite{berg-2007,berg-2008a,berg-2009b,berg-2009}

We end by noting that our results are also applicable to the problem of a two dimensional array of weakly coupled 
spin-1/2 Heisenberg antiferromagnetic spin chains. As shown by Senthil and Fisher~\cite{senthil-2006}, a  super-spin representation used by SF for the $O(4)$ NLSM 
describing N\'{e}el and VBS order parameters for 
this problem is identical to that employed here for us for the problem of coupled stripes. The SC and CDW couplings in the stripe problem correspond to couplings 
between the N\'{e}el and VBS order parameters respectively on nearest neighbor spin chains, while departures from $K_{c}=1$ in our problem of stripes correspond to 
departures from the Heisenberg point to the XXZ model. Thus, our results suggest that in the problem of coupled spin-1/2 chains, there is also a phase in which  
N\'{e}el and columnar VBS orders coexist,
ending at a tetracritical point. Within the coexistence region, the VBS singlet order must have spin triplet excitations. Further, the spatially isotropic $O(4)$ NLSM 
theory of Refs.[\onlinecite{senthil-2004,tanaka-2005}] for the competition of N\'{e}el and VBS orders in the case of the 
2D spin-1/2 Heisenberg antiferromagnetic system admits a deconfined quantum critical fixed point at which 
an anisotropy in the nearest neighbor N\'{e}el and VBS couplings is a relevant operator; the UV stable 
fixed point reached is the anisotropic $O(4)$ NLSM studied here. It is important to note, however, that
our $O(4)$ symmetric theory does not have deconfined topological excitations.
As we have seen, for $(J_{SC},J_{CDW})>0$, the competing orders are 
uniform SC and CDW respectively, while for $(J_{SC},J_{CDW})<0$,the orders are staggered SC (period-2 
pair density wave state) and CDW respectively. In the equivalent spin problem, these two cases 
correspond to the competition between antiferromagnetic N\'{e}el vs. staggered columnar VBS orders and 
ferromagnetic vs. uniform columnar VBS orders respectively.


\begin{acknowledgments}
We thank S. A. Kivelson and T. Senthil for very stimulating discussions.
This work was supported in part by the National Science Foundation, under grant DMR 0758462 at the University of Illinois,
and by the Office of Science, U.S. Department of Energy, under Contract DE-FG02-91ER45439 through the Frederick
Seitz Materials Research Laboratory of the University of Illinois. One of us (S.L.) also thanks NSF DMR~09-06521 
for support.
\end{acknowledgments}

\end{document}